\documentclass[preprint2]{emulateapj-rtx4}

\usepackage{txfonts} 
\usepackage{natbib} 
\usepackage{xspace}
\usepackage{color}
\usepackage[colorlinks,urlcolor=blue,citecolor=blue,linkcolor=blue]{hyperref} 

\newcommand{\methanol}  {CH$_3$OH}

\newcommand{\acetone} {CH$_3$COCH$_3$}
\newcommand{\dimethylether} {CH$_3$OCH$_3$}

\newcommand{\methylformate} {HCOOCH$_3$}

\newcommand{\methylcyanide} {CH$_3$CN}

\newcommand{\ethylcyanide} {CH$_3$CH$_2$CN}

\received{}
\revised{}
\accepted{}

\shorttitle{ALMA observations of the archetypal ``hot core" that isn't: Orion KL}
\shortauthors{Orozco-Aguilera et al.}

\begin{document}

\title{ALMA observations of the archetypal ``hot core" that isn't: Orion KL}

\email{lzapata@crya.unam.mx}

\author{M. T. Orozco-Aguilera}
\affil{Instituto Nacional de Astrof\'\i sica, \'Optica y Electr\'onica, Luis Enrique Erro 1, 
Tonantzintla, Puebla, M\'exico}

\author{Luis A. Zapata}
\affil{Instituto de Radioastronom\'\i a y Astrof\'\i sica,
       UNAM, Apdo. Postal 3-72 (Xangari), 58089\\
       Morelia, Michoac\'an, M\'exico}

\author{Tomoya Hirota}
\affiliation{Mizusawa VLBI Observatory,
                 National Astronomical Observatory of Japan, 
                 Osawa 2-21-1, Mitaka-shi, Tokyo 181-8588, Japan}
                 
\author{Sheng-Li, Qin}                 
\affiliation{Department of Astronomy, Yunnan University, 
                 and Key Laboratory of Astroparticle Physics of Yunnan Province, 
                 Kunming 650091, China 0000-0003-2302-0613}

\author{Masqu\'e, Josep M}
\affil{Departamento de Astronom\'\i a, 
        Universidad de Guanajuato, Apdo. Postal 144, 36000 
        Guanajuato, M\'exico}



\begin{abstract}

We present sensitive high angular resolution ($\sim$ 0.1$''$ -- 0.3$''$) continuum 
ALMA (The Atacama Large Millimeter/Submillimeter Array)  observations 
of the archetypal hot core located in Orion-KL. The observations were made 
in five different spectral bands (bands 3, 6, 7, 8,  and 9) covering a  
very broad range of frequencies (149 -- 658 GHz). Apart of the well-know millimeter 
emitting objects located in this region (Orion Source I and BN),
we report the first submillimeter detection of three compact continuum sources  (ALMA 1-3) 
in the vicinities of the Orion-KL hot molecular core.
These three continuum objects have spectral indices between 1.47 to 1.56, and brightness temperatures between 
100 to 200 K at 658 GHz suggesting that we are seeing moderate optically thick 
dust emission with possible grain growth. However, as these objects are not associated
with warm molecular gas, and some of them are farther out from the molecular core, 
we thus conclude that they cannot heat the molecular core.  
This result favours the hypothesis that the hot molecular core in Orion-KL core is heated externally.       

\end{abstract}




\section{Introduction} 
\label{sec:intro}

Hot Molecular Cores (HMC) are dense ($\ge$ 10$^6$ cm$^{-3}$), warm ($\ge$ 100 K) and compact ($\leq$ 10$^5$ AU) dusty regions within 
molecular clouds that are thought to harbour massive young stars \citep{kur2000}. These regions are characterized by a strong line emission from a large amount 
of molecules, but in particular, from complex organic molecules (COMs), as \methylcyanide, \methanol, HCOOH, \methylformate, 
\dimethylether, \ethylcyanide, and \acetone. COMs are defined by molecules composed with more than six atoms containing C and H elements \citep{her2009}.  
One of the first identified HMC was the one located in the Orion Kleinmann-Low (KL) region \citep{ho1979}. \citet{ho1979} identified 
it as a compact source of hot ammonia emission embedded in a more extended ridge of dense material. Later interferometric 
observations revealed the peculiar and cumply  ``heart" or ``U" morphology of the Orion-KL HMC \citep{gen1982, Wil1994, Wri1996, fav2011}. 
Such morphology is not seen in the southern compact HMCs located in Orion South even observing at the same spatial scales \citep{zap2007,zap2010}.   

The nature of the Orion-KL HMC was first questioned by \citet{bla1996} using continuum/line Owens Valley Radio Observatory (OVRO) observations, 
that found no evidence of a luminous internal heating source within the Orion-KL HMC. This result was more aggravated with Berkeley-Illinois-Maryland Association (BIMA)
 line observations of the formic acid (that also trace the HMC) and the position in the sky of water masers, which suggested that these molecules trace
the interaction region between the outflow and the molecular gas at nearly systemic velocities \citep{liu2002}. 
Moreover, using CO (carbon monoxide)  BIMA observations,  \citet{che1996} proposed that the biconical outflow in Orion-KL is partly 
truncated by the hot molecular core.  

\begin{deluxetable*}{cccccccccc}
\tablecaption{Summary of the observations
 \label{tab-obs}}
\tablehead{
\colhead{}                          & 
\colhead{}                          & \colhead{Center}   & 
\colhead{Total}                     & \colhead{ {\it uv} } & 
\colhead{Number}                          & 
\colhead{On-source} & 
\colhead{Synthesized}                          & \colhead{}    & 
\colhead{}       
\\
\colhead{ALMA}                          & 
\colhead{}                      & \colhead{frequency\tablenotemark{a}}   & 
\colhead{bandwidth}                 & \colhead{range} & 
\colhead{of}                 & 
\colhead{Time} & 
\colhead{Beam}                      & \colhead{PA}    & 
\colhead{rms} 
\\
\colhead{Band}                      & 
\colhead{Date}                 & \colhead{[GHz]}   & 
\colhead{[MHz]}                     & \colhead{[k$\lambda$]} & 
\colhead{Antennas}                  & 
\colhead{[min.]} & 
\colhead{[arcsec]}                  & \colhead{[degree]}    & 
\colhead{[mJy~beam$^{-1}$]}
}
\startdata
4  &       2015/09/04                   &    149.5  &   4000    &     { 7 -- 815}       &      33            &  $\sim$ 8  & 0.36$\times$0.34 &  \ $-$46 & 2.5 \\
4  &       \nodata                         &  \nodata & \nodata &    {150 -- 815}     & \nodata    &  \nodata                & 0.29$\times$0.28 &   \ $-$67 &1.5 \\
6  &       2015/08/28                   &    232.6   &  2342    &  { 200 -- 1480}    &      40   &  $\sim$ 46  & 0.21$\times$0.17 &  \ $+$77 & 4.5 \\
7  &       2014/07/26                   &    348.4  &   7500    &     { 39 -- 870}     &      31            &  $\sim$ 25  & 0.29$\times$0.26 &  \ $-$89 & 6.1 \\
7  &       \nodata                         &  \nodata & \nodata &     { 200 -- 870}    & \nodata    &  \nodata                & 0.26$\times$0.20 &   \ $+$66 &4.6 \\
8  &       2015/09/22                   &    433.7  &  5624  &      { 200 -- 3100}                &      35 &  $\sim$ 25  & 0.10$\times$0.08 &  \ $+$70 & 5.0 \\
9  &       2014/08/05                   &    658.5 &   7500   &    { 300 -- 1400}    &  35 &  $\sim$ 5  & 0.17$\times$0.14 &  \ $-$42 & 34.0 \\
\enddata
\tablenotetext{a}{Center frequency after averaging the four spectral windows.}
\end{deluxetable*}

Recent studies have also confirmed that the HMC in Orion-KL is indeed externally heated possibly by an 
explosive outflow occurred  some 500 yrs ago \citep{zap2009,zap2011a,ball2017} or maybe by the Orion Source I \citep{god2011,wri2017} 
or the expanding bubble-like outflow \citep{zap2011b} . 
Some other  works also confirming this externally heating hypothesis include: \citet{fav2011b,tom2011,peng2012,peng2013,bell2014,gon2015,peng2017,wri2017}.  
For example, \citet{peng2017} found that the Orion-KL HMC emission peaks of vibrationally excited HC$_3$N lines move from south to northeast with 
increasing E$_{u}$, and that the HC$_3$N higher-energy lines have higher rotational temperatures and low column densities which appear to support that the 
hot core is externally heated.  

The strong bursting observed in the water masers located in the direction HMC in Orion-KL could be caused 
by the interaction between the explosive outflow and the ambient quiescent gas \citep{tom2011,tom2014}.  

In this study, using the tremendous sensitivities, a better {\it uv-plane} coverage, and a very broad covering of frequency range at (sub)millimeter wavelengths
 offered by the recent  operational ALMA (The Atacama Large Millimeter/Submillimeter Array)
observatory, we carried out a search for compact and faint continuum sources (in five different continuum bands of ALMA) 
that are probably associated with (proto)stellar objects and with hot molecular gas within the HMC in Orion-KL.  
We report, in addition to Orion Source I and Orion BN, the detection at submillimeter wavelengths of three new compact continuum sources (ALMA1-3) 
that are located in the vicinities of the HMC, but with not hot molecular gas
associated.  Additionally, one of these compact sources seem to be associated with high-mass stars, but it does not account by the internal heating of the HMC. 
We thus conclude that the hot molecular core in Orion-KL is indeed heated externally as suggested by many observational works.          

\section{Observations} 
\label{sec:obs}

The observations were carried out with ALMA between 2014 and 2015 as part of the ALMA programs: 2013.1.01034.S (Band 4), 
2012.1.00146.S (Band 6), 2013.1.00048.S (Band 8), and 2012.1.00123.S (Bands 7 and 9).  The total bandwidth used to estimate the continuum 
emission in the ALMA observations were more than 2 GHz, see Table 1. However, as there are many lines detected in the spectral windows, 
we have probably some contamination from very faint lines. In order to search for compact millimeter sources within the hot molecular 
core in Orion-KL, we constrain the {\it uv-range} of the observations, see Table 1 and Figures 1, 2, and 3. 
We chose the {\it uv-range} based on a trade off between removing as much extended emission as possible and keeping enough visibilities to obtain a good map.
The resulting synthesized beams 
are presented in Table 1.  The number of antennas used during the observations varied between 31 to 40, see Table 1.  
Weather conditions were very good and stable, with an average precipitable water vapour of 6 mm (Band 3),  1 mm (Band 6),  0.3--0.7 mm (Band 7), 
0.3 mm (Band 8), and 0.15 mm (Band 9). 

\begin{figure}[ht!]
\epsscale{1.27}
\plotone{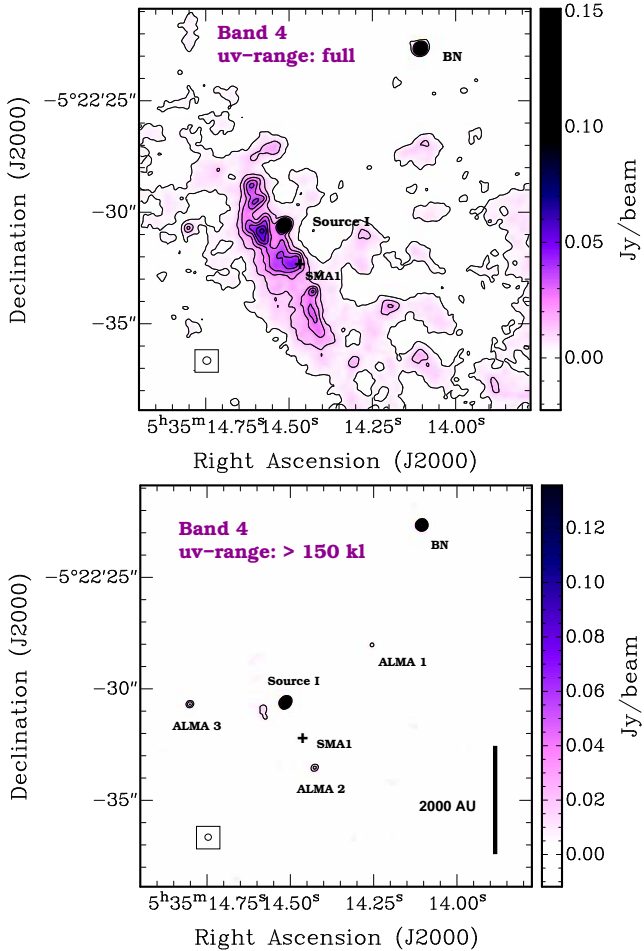} \vspace{-1cm}
\caption{ ALMA continuum band 4 (3.0 mm) contour and colour-scale images of the hot molecular core located in Orion-KL.  
{\bf Upper:} Continuum image using the full {\it uv-range} from the ALMA observations. The contours are starting
from 15$\%$ to 90$\%$ in steps of 5$\%$ of the intensity peak . The intensity peak is 0.150 Jy Beam$^{-1}$. 
The half-power contour of the synthesized beam of the image is shown in the bottom left corner.
{\bf Lower:} Continuum image using only part of the {\it uv-range} ($>$ 150 k$\lambda$) from the continuum observations.
The contours are starting from 15$\%$ to 90$\%$ in steps of 5$\%$ of the intensity peak . The intensity peak is 0.130 Jy Beam$^{-1}$. 
The synthesized beam size is shown in the bottom left corner.
The position of the source SMA1
is indicated with a cross, see \citet{beu2005}. Both images are corrected by the primary beam attenuation.
 {The spatial scale bar is presented in bottom right corner.}}
\end{figure}

\begin{deluxetable*}{ccccccccc}
\tablecaption{Physical parameters of the compact sources$^{a}$}
\tablehead{
\colhead{}     &  \multicolumn{2}{c}{Position$^{b}$}   &\multicolumn{5}{c}{Flux Density} & \colhead{Spectral } \\
\cline{4-8}
\cline{4-8}
\colhead{}     & \colhead{$\alpha_{2000}$} & \colhead{$\delta_{2000}$} &\colhead{Band 4} & \colhead{Band 6}  & \colhead{Band 7} & \colhead{Band 8} & 
\colhead{Band 9} & \colhead{Index} \\
\cline{4-8}
\cline{4-8}
\colhead{Name}    & \colhead{} &   \colhead{}    & \multicolumn{5}{c}{[mJy]} & \colhead{}  
}
\startdata
ALMA 1 & 05 35 14.225  & $-$05 22 28.02 & 18$\pm$2 & 80$\pm$5     & 90$\pm$5    &170$\pm$15 & -- & 1.47$\pm$0.15 \\ 
ALMA 2 & 05 35 14.428  & $-$05 22 33.56 & 41$\pm$5 & 138$\pm$10 & 210$\pm$20 & 305$\pm$25 & 820$\pm$120& 1.48$\pm$0.18 \\
ALMA 3 & 05 35 14.802  & $-$05 22 30.69 & 31$\pm$3 & 117$\pm$15 &  180$\pm$15 &  260$\pm$22 & 720$\pm$60& 1.56$\pm$0.13 \\  
\cline{1-8}
Source I & 05 35 14.515  & $-$05 22 30.60 & 150$\pm$14 & 590$\pm$80 &  905$\pm$90 &  1518$\pm$150 & 3950$\pm$320& 1.74$\pm$0.15\\  
\cline{1-8}
\enddata
\tablenotetext{a}{These parameters were obtained from a Gaussian fit using the viewer in CASA.}
\tablenotetext{b}{Positional errors are $\sim$0.05$''$ for R.A. and decl.}
\end{deluxetable*}

The ALMA calibration includes simultaneous observations of the 183 GHz water line with water vapour radiometers that measure the water 
column in the antenna beam, that is used to reduce the atmospheric phase noise.  Quasars J0529$-0519$, J0423$-$013, J0607$-$0834,  
and J0532$-$0307 were used to calibrate the bandpass, the amplitude and the gain fluctuations.

The data were calibrated, imaged, and analyzed using the Common Astronomy Software Applications \citep[CASA:][]{mac2007}. The data presented 
in this paper were also analyzed using the karma software \citep{goo1996}.  We used ROBUST parameter of CLEAN equal to zero in the continuum maps 
presented in this study.  This was made in order to obtain an optimal compromise between sensitivity and angular resolution.
 The resulting rms-noises for the continuum images and their respective angular resolutions are presented in Table 1. All the resulting ALMA 
 images are corrected by the primary beam attenuation. We self-calibrate all images in phase. { In average, we give about 1 to 2 rounds in phase for
 every observation. This helped to decrease the rms-noises (on a few mJy) and thus increase the S/N ration on 
 the final images. We also tried to do self-calibration in amplitude and phase, but we did not obtained substancial improvements on the images. }

\section{Results} 
\label{sec:res}

\begin{figure}[ht!]
\epsscale{1.27}
\plotone{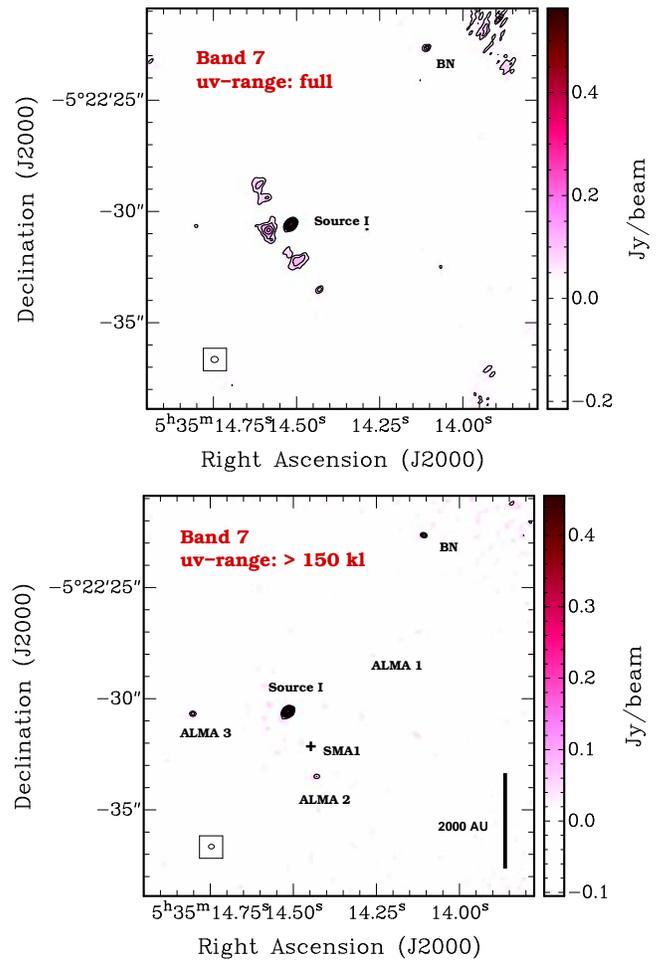} \vspace{-1cm}
\caption{ALMA continuum band 7 (0.8 mm) contour and colour-scale images from the hot molecular core located in Orion-KL.  
{\bf Upper:} Continuum image using the full uv-range from the ALMA observations. The contours are starting
from 30$\%$ to 90$\%$ in steps of 5$\%$ of the intensity peak . The intensity peak is 0.561 Jy Beam$^{-1}$. 
The half-power contour of the synthesized beam of the image is shown in the bottom left corner.
{\bf Lower:} Continuum image using only part of the uv-range ($>$ 150 k$\lambda$) from the continuum observations.
The contours are starting from 10$\%$ to 90$\%$ in steps of 5$\%$ of the intensity peak . The intensity peak is 0.453 Jy Beam$^{-1}$. 
The synthesized beam size is shown in the bottom left corner.
The position of the source SMA1 is indicated with a cross, see  \citet{beu2005}. Both images are corrected by the primary beam attenuation.
 {The spatial scale bar is presented in bottom right corner.}}
\end{figure}
   
\begin{figure}[ht!]
\epsscale{1.1}
\plotone{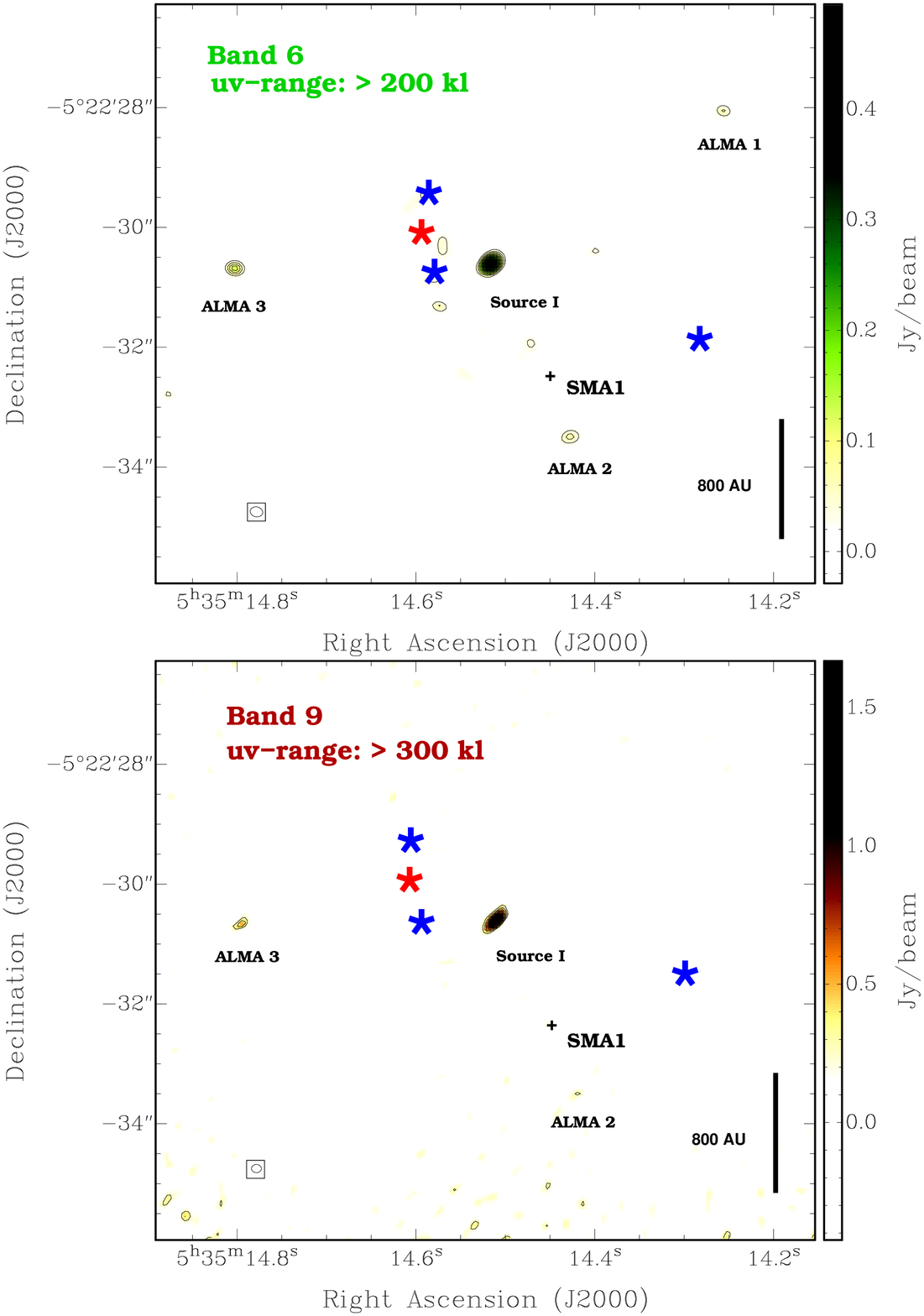}
\caption{ ALMA continuum bands 6 and 9 contour and colour-scale images of the hot molecular core located in Orion-KL.  
{\bf Upper:} Band 6 continuum image using part uv-range from the ALMA observations ($>$ 200 k$\lambda$). The contours are starting
from 30$\%$ to 90$\%$ in steps of 5$\%$ of the intensity peak . The intensity peak is 0.461 Jy Beam$^{-1}$. 
The half-power contour of the synthesized beam of the image is shown in the bottom left corner.
{\bf Lower:} Band 6 continuum image using part uv-range from the ALMA observations ($>$ 200 k$\lambda$). 
The contours are starting from 55$\%$ to 90$\%$ in steps of 5$\%$ of the intensity peak . The intensity peak is 1.533 Jy Beam$^{-1}$. 
The first contour level, in the lower panel is about 10-$\sigma$ (45 mJy Beam$^{-1}$).
The half-power contour of the synthesized beam of the image is shown in the bottom left corner. The position of the source SMA1
is indicated with a cross, see \citet{beu2005}. Both images are corrected by the primary beam attenuation. The blue asterisks mark
the position of the peaks of the hot molecular gas as traced by the NH$_3$(12,12) with an upper level energy over the ground of 1456 K, 
see  \citet{god2011}.  The red asterisk marks the position of the peak of the hot molecular gas as traced by the CH$_3$CN(12$_{9}$,11$_{9}$) 
with an upper level energy over the ground of 646 K, see  \citet{zap2011b}. {The spatial scale bar is presented in bottom right corner of each panel.}}
\end{figure}

 \begin{figure}[ht!]
\epsscale{1.25}
\plotone{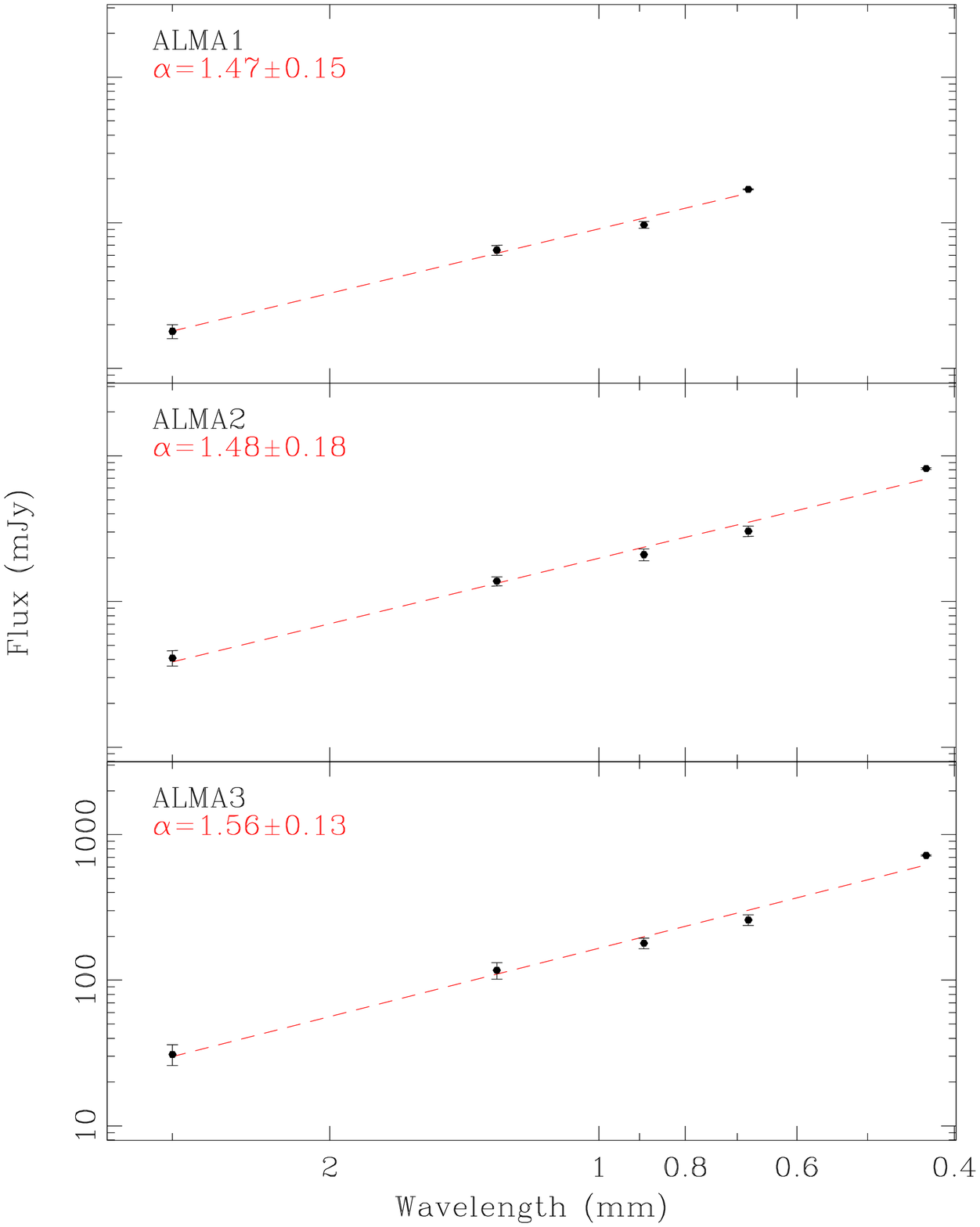} 
\caption{{Spectral energy distributions (SEDs) of the detected compact sources from the millimeter to submillimeter wavelengths. 
                    The dashed lines are a least-squares power-law fit (of the form S$_\nu$ $\propto$ $\nu^\alpha$ to the spectrum). The $\alpha$-values 
                    of the fitting are shown in the panels.
}}
\end{figure} 

The main results of this study are presented in Table 2, and Figures 1, 2, and 3. 

In Figure 1, we present the results of the ALMA observations 
in band 4 (149.5 GHz). In this Figure, we include the observations of the HMC in Orion-KL using the full {\it uv-range} (see Table 1)
and one map only using baselines that are sensitive to compact emission ($>$ 150 k$\lambda$) or structures with angular sizes of less than 1.3$''$.    
In the image using the full {\it uv-range} (upper panel), the HMC in Orion-KL is well detected together with the two strong compact objects (Orion Source I and BN) at
these millimeter wavelengths. The ALMA band 4 observations revealed the ``heart" or ``U" shape of the HMC as already traced by many authors, see
for example \citet{god2011}, where it is mapped by the NH$_3$ line emission.  This ``heart" structure is very well resolved, with its peculiar clumpy structure. On the other hand,
the map with the restricted {\it uv-range}, revealed 6 compact sources above the 5-$\sigma$ (7.5 mJy Beam$^{-1}$) noise level, see the lower panel of Figure 1. However, as some of these
compact sources are not detected in other ALMA bands, we only count five real detections (Orion Source I, BN, ALMA1, ALMA2, and ALMA 3).
We give the physical parameters of these new sources in Table 2.  We note that the submillimeter source SMA1 found using Submillimeter Array 
(SMA) observations and reported by \citet{beu2005} does not have any counterpart, perhaps it is resolved out by the present ALMA observations.

In Figure 2,  we present the results of the ALMA observations in band 7 (348.4 GHz).
In this Figure, we include the observations of the HMC using the full {\it uv-range} (see Table 1)
and one map only using baselines that are sensitive to compact emission ($>$ 150 k$\lambda$) or 
structures with angular sizes of less than 1.3$''$.  In the image using the full {\it uv-range} (upper panel), only the most compact structures of the HMC 
are detected. These compact and clumpy structures are located close to Orion Source I and form an arc-like structure. 
On the other hand, the map with the restricted {\it uv-range}, revealed 4 compact sources above the 5-$\sigma$ (23 mJy Beam$^{-1}$) 
noise level (see the lower panel of figure 2). ALMA 1 is present in the map, even though its emission is very faint. 
Despite starting the contours at 10-$\sigma$, we consider all sources below this level are spurious because they do not show 
well-defined counterparts in all other ALMA bands, contrary to the objects ALMA1-3.
We give the flux densities at this ALMA band and from these new sources in Table 2.

In Figure 3, we present the results of the ALMA observations in band 6 and 9 (232.6 and 658.5 GHz). 
The data from ALMA band 8 is very similar to that obtained in band 7 and it is not shown in this study.
In Figure 3, we only include the observations of the HMC using baselines that are sensitive to compact emission ($>$ 200 k$\lambda$ for band 6,
 and $>$ 300 k$\lambda$ for band 6) or structures with angular sizes of less than 1$''$ and 0.65$''$, respectively.
 In these maps,  the Orion BN object is far from the phase center and it is very attenuated, so that it is not shown in the images of Figure 3. However, we 
 see clearly the rest of the objects (Orion Source I, ALMA1, ALMA2, and ALMA3). In this image, we have
 additionally included the positions of the peaks of the hot molecular gas as traced by the NH$_3$(12,12) 
 with a temperature in the upper level of 1456 K, see  \citet{god2011}.  The positions of the peak of the hot molecular gas as traced 
 by the CH$_3$CN(12$_{9}$,11$_{9}$) with a temperature in the upper level of 646 K is also included 
 in Figure 3, see  \citet{zap2011b}. None of these positions coincide with the submillimeter compact sources
 reported in this paper.  ALMA 1 is not detected in the band 9 observations because, possibly, the observations are too noisy and, 
 in addition, this source is close to the edge of the primary beam.  
 
 {In Figure 4, we present the spectral energy distributions (SEDs) of the detected compact sources from the millimeter
  to submillimeter wavelengths. }

\section{Discussion and Conclusions} 
\label{sec:con}

A discussion on the nature of the strong sources in the field, Orion Source I and
BN at submillimeter wavelengths is outside of the scope of the paper, but we refer to the readers 
to the papers of \citet{beu2006,tom2015,tom2016,plam2016}, where a complete discussion is made
using ALMA and SMA data.

As we have observations from 149.5 GHz up to 658.5 GHz of the Orion-KL HMC, we can construct the spectral 
indices (S$_\nu$= $\nu^\alpha$, where S$_\nu$ is the density flux at different frequencies, $\nu$ is the frequency and $\alpha$, 
the spectral index) of the compact objects by use of the new ALMA multiple band observations ({Figure 4}). We estimated the values of the spectral indices
using the least-squares power-law technique. 
The estimated values of spectral indices are presented in Table 2. 
They are relatively not so steep compared with those reported to be associated with optically thick thermal dust emission (about 2 -- 3), but
probably it is due to possible grain growth that flattens the spectral indices \citep{dra2006}. We thus conclude that possibly we are detecting 
moderate optically thick thermal dust emission.  Taking the values of their flux densities at 658 GHz, we estimated brightness temperatures
of about 100 to 200 K.  This again suggests that we are seeing dust emission.  An estimation of the mass from the dust emission for these submillimeter 
objects seems not to be a reliable as the emission is moderate optically thick.   Finally, we also construct the spectral index  for Source I and compared
the resulting value (see Table 2) with those reported in the literature, see \citet{tom2015,tom2016,plam2016}. The value of 1.8 for spectral index for
 Source I reported in  \citet{tom2015} is the closest one to that found in this study, which confirm that the derived spectral indices of the 
 ALMA1-3 continuum sources in this work are reliable. { We also have compared the flux densities of the ALMA compact sources reported here with
 the full and restricted uv-ranges, and find that these are very similar.  Moreover, we restricted the images to similar uv-ranges for band 6 and 9 observations, 
 and find not different values for the flux densities presented in Table 2.}

We found that ALMA 1 is well coincident (within a 0.8$''$ error) with
the millimeter source found in CARMA observations, FW2011-C14 \citep{fri2011}, and the mid-infrared extended object called [RLK73] IRc6E  \citep{shu2004}.
However, as the mid-infrared is very extended it is difficult to know if they are really connected. For ALMA 2 and ALMA 3, we also find that, within the position errors, these
submillimeter objects are coincident with the CARMA millimeter objects  FW2011-C14 and FW2011-C22 \citep{fri2011}, respectively. 
We found that ALMA 3 additionally is associated with the compact mid-infrared object  [RLK73] IRc12 \citep{shu2004}.
 \citet{rob2005} reported that [RLK73] IRc12 is a very luminous object with a bolometric luminosity of 4$\times$10$^4$ L$_\odot$,
 between 7.7 and 12.4 $\mu$m. This luminosity corresponds to a high-mass star.

As mentioned before in Figure 3, we have included the positions of the peaks of the hot molecular gas as traced by the NH$_3$(12,12)
and  CH$_3$CN(12$_{9}$,11$_{9}$), and found that these positions do not coincide with any of the submillimeter objects reported in
this study.  Even the infrared object [RLK73] IRc12 associated with ALMA3 and that is maybe a high-mass (proto)star, is too far away 
from the hot molecular gas. Additionally the other IR sources and massive stars are also not associated with hot molecular gas.   
We thus concluded that the hot molecular gas in the Orion-KL core probably is heated externally. 

If the HMC is heated internally, we expect young massive stars in middle of hot molecular gas, but if the heating 
is externally the core should be illuminated from the edges, this physical effect has been already traced by radio 
observations made by \citet{zap2011b,god2011}.  According to the results reported in the present paper, and 
those from \citet{zap2011b,god2011} an externally heating model seems to fit much better with the HMC in Orion-KL.
We thus favour the hypothesis that the hot molecular core in Orion-KL core is heated externally.


\acknowledgments

This research has made use of the SIMBAD database, operated at CDS, Strasbourg, France. 
L.A.Z. is grateful to CONACyT, Mexico, and DGAPA, UNAM for their financial support. 
S.-L. Q. is supported by NSFC under grant No. 11373026, and Top Talents Program of Yunnan
Province (2015HA030). T.H. is supported by MEXT/JSPS KAKENHI grant Nos. 21224002, 24684011,
and 25108005, and the ALMA Japan Research Grant of NAOJ Chile Observatory, NAOJ-ALMA-0006.
We are very thankful for the thoughtful suggestions of the
anonymous referee that helped to improve our manuscript.
This paper makes use of the following ALMA data: \\

 \noindent ADS/JAO.ALMA\#2012.1.00146.S, 
 ADS/JAO.ALMA\#2013.1.00048.S, 
 ADS/JAO.ALMA\#2013.1.01034.S, 
 ADS/JAO.ALMA\#2012.1.00123.S.\\

ALMA is a partnership of ESO (representing its member states), NSF (USA) and NINS (Japan), 
together with NRC (Canada) and NSC and ASIAA (Taiwan) and KASI (Republic of Korea), 
in cooperation with the Republic of Chile. The Joint ALMA Observatory is operated by ESO, 
AUI/NRAO and NAOJ.




\end{document}